\documentstyle[psfig]{elsart}

\begin{document}

\begin{frontmatter}

\title{Deuteron photo-disintegration with polarised photons in the energy
range 30 - 50 MeV}

\author[1]{D. Babusci}, 
\author[2,3]{V. Bellini}, 
\author[4,5]{M. Capogni}, 
\author[4]{L. Casano}, 
\author[4]{A. D'Angelo}, 
\author[6,7]{F. Ghio}, 
\author[6,7]{B. Girolami}
\author[4,8]{D. Moricciani}, 
\author[4,5]{C. Schaerf}

\address[1]{INFN - LNF, {\it Via Enrico Fermi 40, I-00044 Frascati, Italy}}
\address[2]{Physics Department of the University of Catania, {\it Corso Italia
57, I-95129 Catania, Italy}}
\address[3]{INFN - LNS, {\it Viale S. Sofia 44, I-95125 Catania, Italy}}
\address[4]{INFN - Roma2, {\it Via della Ricerca Scientifica 1, I-00133 
Rome, Italy}}
\address[5]{Physics Department of the University of Rome "Tor Vergata", {\it 
Via della Ricerca Scientifica 1, I-00133 Rome, Italy}}
\address[6]{Istituto Superiore di Sanit\`a, {\it Viale Regina Elena 299, 
I-00161 Rome, Italy}}
\address[7]{INFN - ISS, {\it Viale Regina Elena 299, I-00161 Rome, Italy}}
\address[8]{Present Address: ISN, {\it Avenue des Martyrs 53, F-38026 Grenoble 
Cedex, France}}

\begin{abstract}
The reaction $d(\vec\gamma,np)$ has been studied using the  tagged and 
polarised LADON gamma ray beam at an energy $30 - 50~MeV$  
to investigate the existence of narrow dibaryonic resonances 
recently suggested from the experimental measurements in a different 
laboratory. 
The beam was obtained by Compton back-scattering of laser light on the 
electrons of the storage ring ADONE. Photo-neutron yields were measured 
at five neutron angle $\vartheta_n^{c.m.} = 22^\circ$, $55.5^\circ$, 
$90^\circ$, $125^\circ$ and $157^\circ$ in the center of mass system.
Our results do not support the existence of such resonances.
\end{abstract}

\end{frontmatter}

\centerline{PACS: 25.20 D, 24.70, 29.27 H}

\section{Introduction}
The deuteron is the most fundamental nuclear laboratory and for this reason,
deuteron photo-disintegration $d(\vec\gamma,np)$ has been extensively studied for 
the last sixty years \cite{chad}. 
The center of mass (c.m.) differential cross section for the reaction 
$d(\vec\gamma,n)p$ induced by linearly polarised gamma rays has been 
calculated, measured and discussed by several authors. We refer here 
to some extensive contributions only \cite{arh,partovi,cmr}. It 
can be written in the form:

\begin{eqnarray}
\frac{d\sigma}{d\Omega}(E_\gamma,\vartheta_n,\varphi_n)&=&I_0(E_\gamma,
\vartheta_n)+PI_1(E_\gamma,\vartheta_n)\cos 2\varphi_n\nonumber \\
&=&I_0(E_\gamma,\vartheta_n)\biggl\lbrace 1+P\Sigma(E_\gamma,\vartheta_n)\cos 
2\varphi_n\biggr\rbrace
\label{ds_deu}
\end{eqnarray}

\noindent where $\vartheta_n$ is the angle between the neutron and photon 
momentum in the c.m. system and $\varphi_n$ is the angle between the direction 
of the polarisation of the incoming photon and the reaction plane; $P$ represents 
the degree of linear polarisation of the photon beam.  The expression 
(\ref{ds_deu}) is obtained only taking into account the spin of the photon.
The microscopical structure of the $N-N$ interaction or the contribution
of meson exchange currents (MEC) or the internal excitation of photo-disintegration 
configurations (IC), or the presence of sub-nucleonic degrees of freedom, 
influence only the form of the functions $I_0(E_\gamma,\vartheta_n)$ and 
$I_1(E_\gamma,\vartheta_n)$. 
\par
Recent experimental measurements of the deuteron photo-disintegration cross 
section made in Kharkov \cite{russi} with a linearly polarised photon beam 
obtained by coherent bremsstrahlung have shown some evidence of three narrow 
dibaryonic resonances in the differential cross section in the plane 
perpendicular to the beam polarisation $\varphi_n=90^\circ$: 

\begin{eqnarray}
\frac{d\sigma}{d\Omega}(E_\gamma,\vartheta_n=90^\circ,\varphi_n=90^
\circ)=I_0(E_\gamma,\vartheta_n=90^\circ)-I_1(E_\gamma,\vartheta_n=90^\circ)
\end{eqnarray}

\noindent they appeared at a very low excitation energies of the ({\it n$-$p}) 
system as indicated in table 1 where we have shown the total energy of the 
({\it n$-$p}) system, the corresponding gamma ray energy and the apparent 
width: 

\par
\vspace{0.6cm}
\centerline{\bf Table 1}
\medskip
\centerline{\begin{tabular}{||c|c|c||}
\hline
 $E_{(n-p)}$  & $E_\gamma$ & $\Gamma$   \\
  $(MeV)$     &  $(MeV)$   &  $(MeV)$   \\ \hline
   1919.5     &   43.9     &    4.5     \\
   1933       &   57.4     &    2.7     \\
   1942       &   66.4     &    6.6     \\ \hline
\end{tabular}}
\par
\medskip
\vspace{0.6cm}

\begin{figure}
\vspace{4.5truein}
\includegraphics{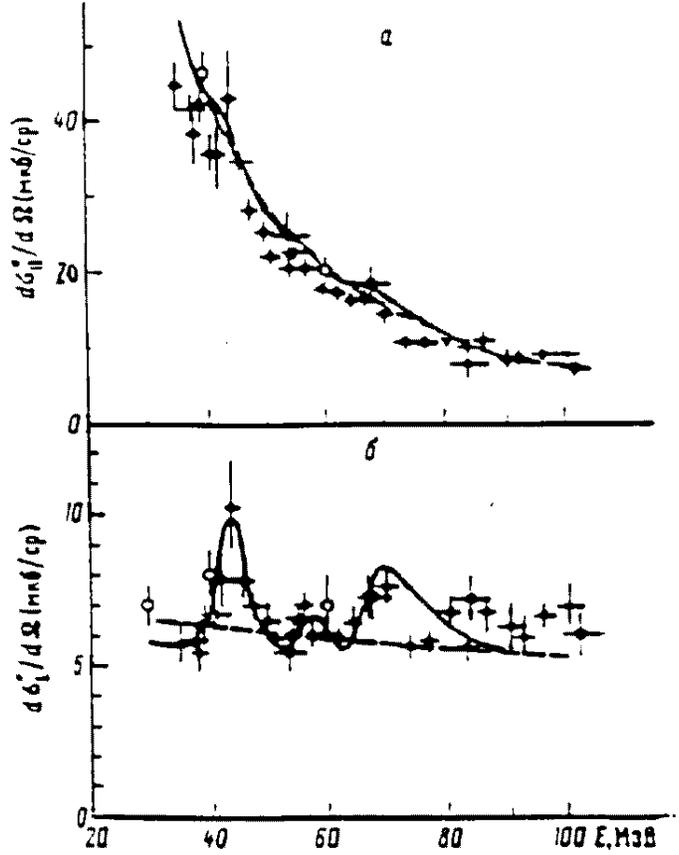}
\caption{$\Bigl (\frac{d\sigma}{d\Omega}\Bigr )_{\parallel,\perp}(E_\gamma,
\vartheta_n=90^\circ)$ from reference [5]; the symbol $\circ$ 
shows the previous data obtained with the LADON beam [6]}
\label{dib_russi}
\end{figure}

In figure \ref{dib_russi} are shown the results obtained at Kharkov compared 
with the previous results obtained with the LADON beam \cite{ladon_ab}. 
The amazing result is that the Kharkov data are in good agreement with ours 
since their resonances are located between our points.

To verify the existence of the first and most impressive of these resonances 
we have taken advantage of the complete polarisation ($P\sim 1$) and good 
energy resolution of our tagged LADON beam $\sigma_{E_\gamma}\simeq 2~MeV$ 
\cite{taladon,ladon1}.

\section{Experimental Setup}

\subsection{The LADON $\vec\gamma$ beam}

One of the most interesting features of the Compton backscattered $\vec\gamma$ 
ray beam is its polarisation: the $\vec\gamma$ rays emitted in the backward 
direction with energy close to the maximum have the same polarisation 
of the initial laser photons.
Data were collected at different $\vec\gamma$ ray energies obtained by 
changing the incoming electron energy. The energy of the $\vec\gamma$ beam 
has been measured with an internal tagging detector where the scattered 
electrons are momentum analysed by one dipole and one quadrupole magnet of 
the ADONE storage ring lattice. The tagging system consists of a silicon 
solid-state $\mu-$strip detector composed of 96 vertical strips with a pitch 
of $650~\mu m$, backed by a fast plastic scintillator. 
\par
The energy resolution of the $\vec\gamma$ beam depends on the energy of the
scattered electron and the energy of the electrons circulating in the storage 
ring, but in any case we obtain $\sigma_{E_\gamma}\le 2.2~MeV$ 
\cite{ladon1}.
During this experiment data are collected using different maximum energy 
of the $\vec\gamma$ beam $E_\gamma^{max}=35,~38,~41,~45$ and $50~MeV$ in 
order to scan with high accuracy the energy region where the first dibaryonic
resonance is proposed.
\par
The photon flux as been measured with a cylindrical $NaI$ detector of $25.4~cm$ 
length and $25.4~cm$ diameter with an efficiency for photon detection of 
$\sim 100~\%$.

\subsection{Detectors}

The target cell is an aluminium cylinder with a diameter of $3.81~cm$ full with 
deuterated liquid scintillator NE230 (full target) made of $C_6D_6$. 
When a deuteron in the target disintegrates the proton does not have enough 
energy to leave the target and deposits all its energy into the target. The 
target is viewed by a photo-multiplier which provides a 
signal proportional to the energy deposited by the proton. The energy 
threshold used for this detector is $3~MeV$, while the minimum energy for 
the proton coming from the deuteron photo-disintegration in our experimental 
condition is $12~MeV$. For this reason we can reasonably assume a proton 
detection efficiency of $\sim 100\%$.
\par
The neutrons escape the target and are detected by five time of flight (TOF) 
detectors made of horizontal cylinders, $30.4~cm$ of diameter and 
$15.4~cm$ of length,  filled by organic liquid scintillator, NE213. 
These detectors are placed at a distance $D\simeq 60~cm$ from the target 
and at angles $\vartheta_n=22^\circ,~55.5^\circ,$ $90^\circ,~125^\circ$ 
and $157^\circ$. Each of them covers a solid angle of $0.13~sr$. The threshold
on the amplitude of the signal from these detectors is $0.5~MeV$. The TOF 
between the proton pulse in the target and the neutron pulses are obtained with a 
resolution (FWHM) of $\Delta T\simeq 1.3~ns$ (this value has been measured 
with the coincidence of the two $\gamma$ photons emitted by a $^{60}$Co 
source). 
Comparing the TOF of the neutrons (coming from the photo-nuclear reaction on
the target) with that of the $\gamma$ (Compton scattered in the target)
detected in these counters we were able to have a reasonable measurement of 
the energy of the neutron by its TOF and to discriminate the neutrons against 
the e.m. background produced in the target.
Calling $\widetilde t = t_n - t_\gamma$ the TOF difference between a neutron 
and $\gamma$, the kinetic energy of the neutron is given by the following 
relation: 

\begin{eqnarray}
T_n=M_n c^2\Biggl (\sqrt{1+\frac{D^2}{c^2 \widetilde t^2+2Dc \widetilde t}}-1
\Biggr ).
\label{T_n}
\end{eqnarray}

To estimate the contribution of the background of events coming from the aluminium 
walls of the cell or the Carbon present in the full target we have also 
taken data with a second target (empty target) NE231 made of $C_6H_6$, 
similar to the first one but with hydrogen instead of deuterium. 
\par
\begin{figure}
\centerline{\psfig{file=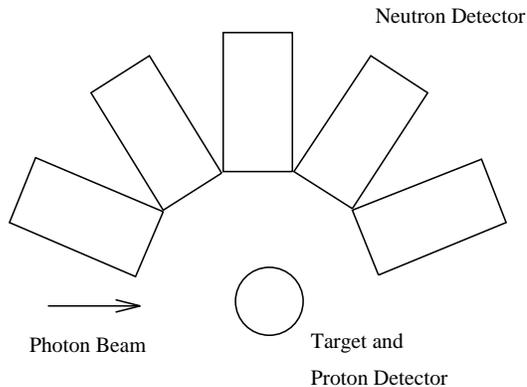,height=5cm}}
\caption{Experimental apparatus}
\label{apparato}
\end{figure}

The experimental apparatus is shown in figure \ref{apparato}. Using this 
apparatus we have measured in coincidence the distribution of the protons 
pulses and the TOF of the neutrons emitted in the photo-reaction on the 
target.

\section{Data Analysis}

The data analysis consists of two steps:

\begin{itemize}
\item[1)] calibration of the apparatus, i.e. the tagging system, the active
target for the proton detection, the neutron detectors and the estimation of 
their efficiencies;
\item[2)] selection of the $d(\vec\gamma,np)$ events and the calculation of 
the differential cross section;
\end{itemize}

The tagging system was calibrated using a magnetic pair spectrometer in 
coincidence, in our energy range the tagging energy calibration is 
linear as illustrated in \cite{taladon,ladon1} and its efficiency is 
$\varepsilon_{tag}=0.96\pm 0.03$ . 
\par
\begin{figure}
\centerline{\hbox{
\psfig{file=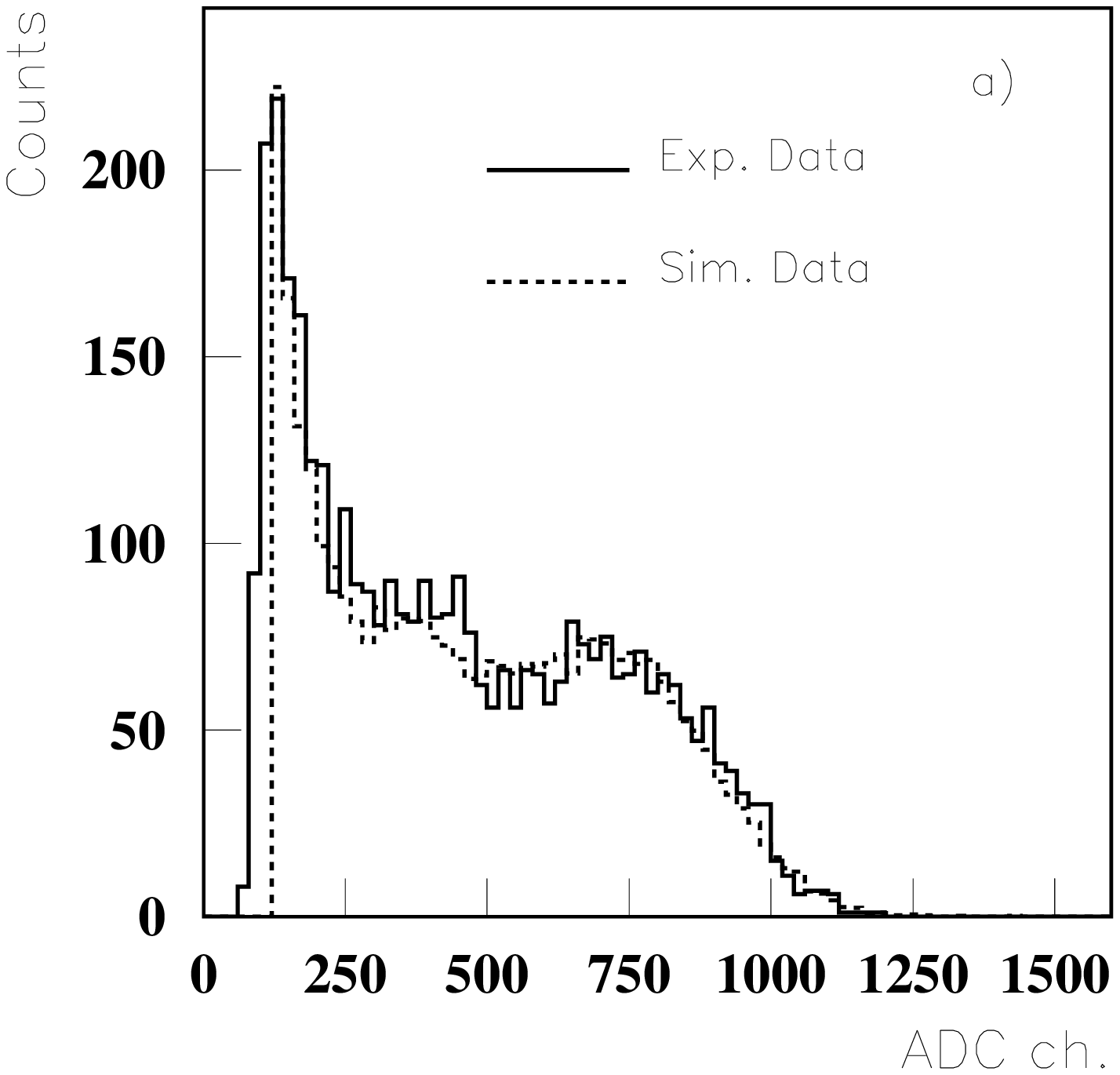,height=7cm,width=7cm}
\psfig{file=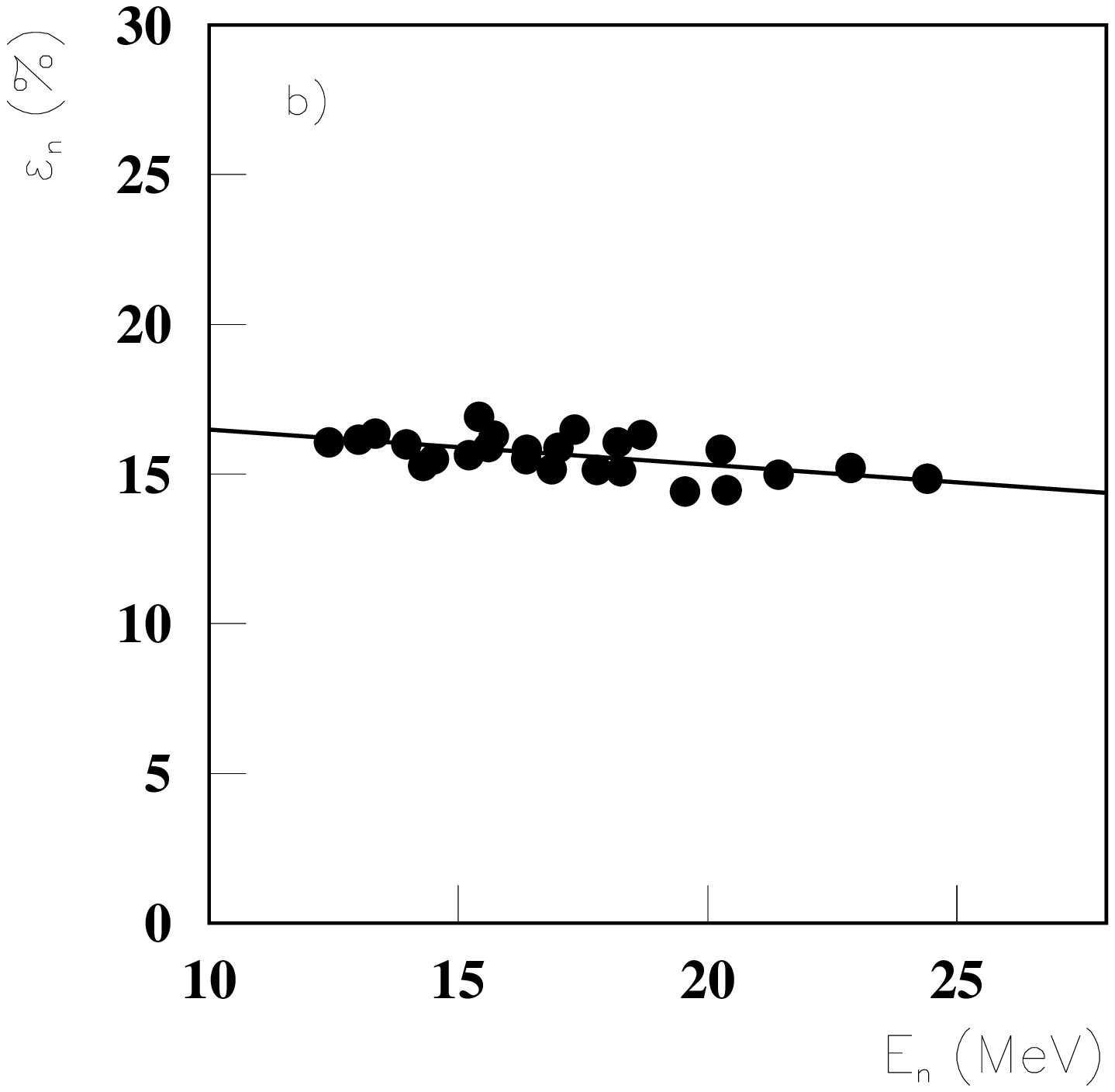,height=7cm,width=7cm}
}}
\caption{a) Comparison between the experimental ADC spectra associated with the neutron 
detection and the ones simulated, b) Neutron efficiency as function of neutron 
energy}
\label{eff_n}
\end{figure}

The TOF calibration of the neutron detectors is determined by reference to 
the $\gamma$ Compton diffused by the electrons in the target.
The neutron detection efficiency was calculated using a Monte Carlo code
\cite{anghi} which takes into account all the nuclear reactions 
on the proton and carbon of the NE213 scintillators by the neutron 
coming from the deuteron photo-disintegration \cite{hess,McLane,DalGuerra,neu3},
and also experimental effects produced by the electronic chain, associated 
with each detectors, and the effect of the threshold used. 
A comparison between the experimental ADC spectra and the simulated one is 
shown in figure \ref{eff_n}a), from this figure is derived the neutron 
detection efficiency which is quite constant as function of the neutron 
energy, as illustrated in figure \ref{eff_n}b). 
From this we obtain its average value and its error 
$\langle\varepsilon_n(E_n)\rangle = (15.6\pm 0.1) \%$.

The energy calibration of the active target is made using the kinematics of   
deuteron photo-disintegration, for which we know the energy of the incoming 
photon and the angle and energy of the outgoing neutron (in the LAB. system). 
We can use this information to determine the kinetic energy of the proton and
comparing it with the ADC of the active target we derive its calibration.
\par
The second step of the analysis which consists on the selection of the $d(
\vec\gamma,np)$ events is also divided in two steps:

\begin{itemize}
\item[1)] the rejection of the e.m. events due to Compton scattering in 
the target;
\item[2)] the identification of the $d(\vec\gamma,np)$ events, respect to other 
nuclear reaction coming from different reactions.
\end{itemize}

\begin{figure}
\centerline{\psfig{file=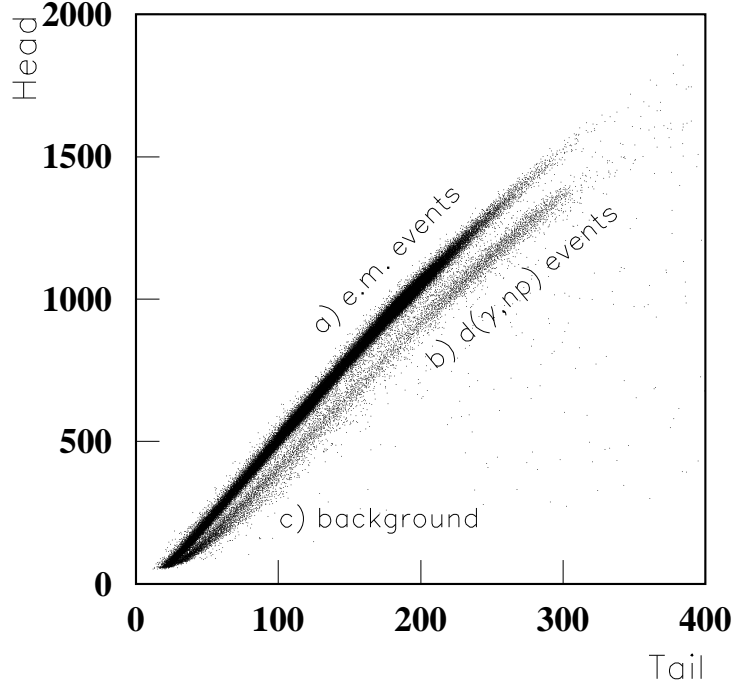,height=9cm}}
\caption{Example of Head-Tail scatter plot from the active target data. For 
goods events the Head values must be proportional to the energy of proton 
emitted in the deuteron photo-disintegration. It must be greater then a 
given value which depend on the kinematics of the event and is around 
channel $700$.}
\label{testa-coda}
\end{figure}

The rejection of the e.m. background was facilitated by the fact that in our 
target the form of the pulse associated with the detection of a hadron is 
different from that of an electron/photon, thus we have used the Head-Tail 
\cite{sj,birks,owen,kl,brooks} technique. The Head is defined as the 
integral of the entire pulse, which is the signal which provides the energy 
of proton, while the Tail is the integral of the last part of it. 
This procedure allowed the separation of the nuclear events as shown in figure 
\ref{testa-coda}, where it is possible to recognise three types of events: 
$a)$ the e.m. events, $b)$ the nuclear events coming from the deuteron 
photo-disintegration and $c)$ the nuclear events coming from photo-reaction on 
target walls or in the carbon also present in the target. The nuclear events
$b)$ and $c)$ are clearly separated from those of type $a)$ and we introduce 
a graphical cut on this plot to isolate them.

Further the identification of the nuclear events coming from reaction under
study was done by a minimisation of the following variable: 

\begin{eqnarray}
\chi^2(E_\gamma,\vartheta_n,E_n,E_p)&=&
\frac{(E_\gamma^{meas}-E_\gamma^{theo})^2}{\sigma_{E_\gamma}^2}+
\frac{(\vartheta_{n}^{meas}-\vartheta_n^{theo})^2}{\sigma_{\vartheta_n}^2}\
\nonumber \\
&+&\frac{(E_n^{meas}-E_n^{theo})^2}{\sigma_{E_n}^2} 
+\frac{(E_p^{meas}-E_p^{theo})^2}{\sigma_{E_p}^2}
\label{chi2}
\end{eqnarray}

\noindent where the quantities with the superscript $^{meas}$ are those 
experimentally measured while the quantities with the superscript $^{theo}$ 
are calculated using the conservation of energy and momentum in our two body 
reaction. $\sigma_{E_\gamma}$, $\sigma_{\vartheta_n}$, $\sigma_{E_n}$ and 
$\sigma_{E_p}$ are the uncertainties in the experimental quantities and are 
know with a error less than $20\%$. The reaction $d(\vec\gamma,np)$ is a two 
body one and its kinematics is completely determined if only the energy of the
incoming photon $E_\gamma$ and the angle of the outgoing neutron $\vartheta_n$
in the LAB. are known.
For this reason in the minimisation procedure we use as independent 
variables these two, varying $E_\gamma$ in the interval $(E_\gamma^
{meas}-3\sigma_{E_\gamma},E_\gamma^{meas}+3\sigma_{E_\gamma})$ and 
$\vartheta_n$ inside the solid angle covered by the neutron detector. 
The selection of the $d(\vec\gamma,np)$ events can now be done using a cut 
in a combination of the dependent variables of the right-hand side of 
equation (\ref{chi2}).
\par
\begin{figure}
\centerline{\psfig{file=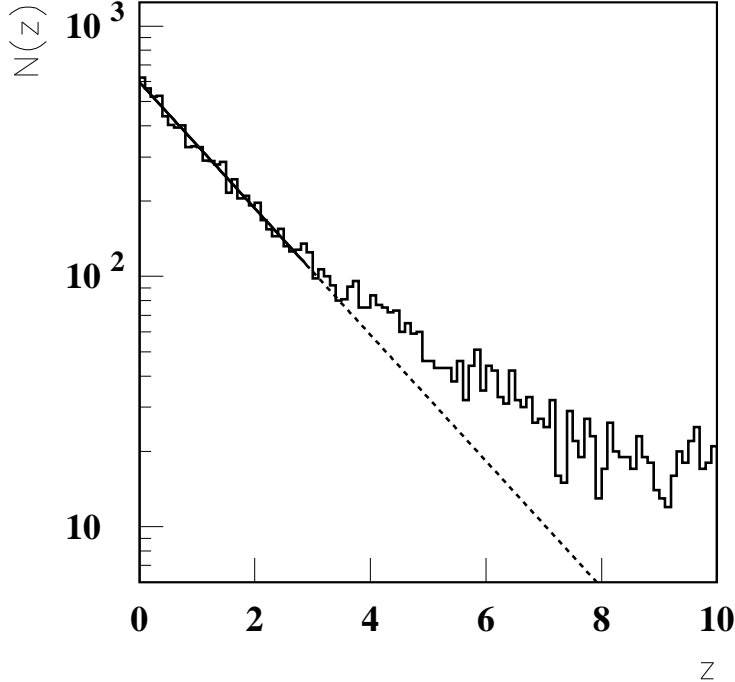,height=9cm}}
\caption{Distribution of $z$ for the full target events. The full line is 
a fit using the function $A_\chi\chi^2(z;2)$ for $z<3$ while the dashed line 
is its extrapolation for $z\in (3,10)$.}
\label{z}
\end{figure}

For each event we define the new variables $x_n=E_n^{theo}-E_n^{meas}$ and
$x_p=E_p^{theo}-E_p^{meas}$. The distribution of $N(x_n,x_p)$ can be fitted by 
the following expression: 

\begin{eqnarray}
N(x_p,x_n)=Ae^{-\frac{(x_p-\mu_p)^2}{2\eta_p^2}} e^{-\frac{(x_n-\mu_n)^2}
{2\eta_n^2}},
\label{npn}
\end{eqnarray}

\noindent and using the parameters $A$, $\mu_p$, $\eta_p$, $\mu_n$ and $\eta_n$ 
it is now possible to define a new variable: $z=\frac{(x_p-\mu_p)^2}{\eta_p^2}+
\frac{(x_n-\mu_n)^2}{\eta_n^2}$ which follows a  $\chi^2$  distribution 
with two degrees of freedom: $\chi^2(z;2)=\frac{exp(-z/2)}{2}$. The 
experimental distribution of $z$ is shown in figure \ref{z}.

The excellent agreement between our data and the expression $A_\chi \chi^2(z;
2)$ for $z\le z_{max}\simeq 3$ confirms the validity of this procedure. 
Studying the distribution of $z$ for the three groups of events indicated in 
figure \ref{testa-coda} we clearly see that the e.m. events coming from the 
class $a)$ have $z>\sim 10$, the events coming from class $b)$ have $z<\sim 
3 - 3.5$ while the events of the class $c)$ have $z>\sim 4$. In conclusion the 
number of events of deuteron photo-disintegration is given by the following 
relation:

\begin{eqnarray}
N_{ev}=N_{ev}(z\le z_{max})+A_\chi\int_{z_{max}}^\infty\chi^2(z;2)dz.
\label{n_ev}
\end{eqnarray}

\begin{figure}
\centerline{\psfig{file=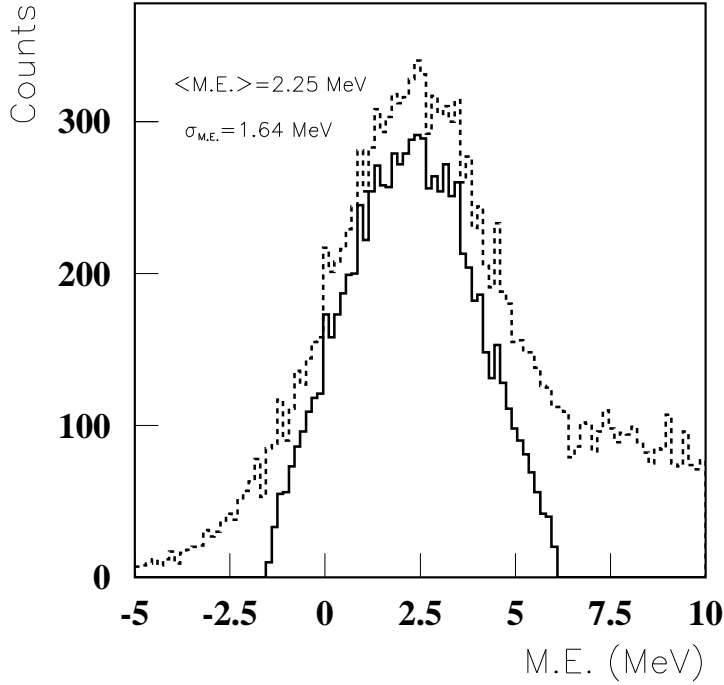,height=9cm}}
\caption{Missing Energy for the reaction $d(\vec\gamma,np)$, dashed line 
all nuclear events(class $b$ and $c$ from the figure \ref{testa-coda}), 
full line events with the selection $z<3$.}
\label{missm_np}
\end{figure}

Where the first term is the number of events clearly identified while the 
second is an estimation of the good events mixed with the backgrounds. The
second term is typically $2\div 3\%$ of the first suggesting a systematic error 
in the estimate of the cross section of the order $\sim 1\%$. In figure 
\ref{missm_np} we have plotted the missing energy, $M.E. = E_\gamma-E_p-E_n$, 
in our reaction. For the events identified with this procedure ($z<\sim 3$)
its average value is very close to the deuteron binding energy $2.2~MeV$
and the resolution is a few $MeV$. Its confirms that the result of our procedure 
is correct. 

\section{Experimental results and conclusion}  

The experimental cross section has been calculated according to:

\begin{eqnarray}
\frac{d\sigma}{d\Omega}=\frac{N_{ev}}{N_\gamma~L~N_d~\Delta\Omega_n~
\varepsilon_n}
\label{cros}
\end{eqnarray}

\noindent where $N_{ev}$ represent the nuclear events calculated with the 
procedure illustrated in the previous paragraph, $N_\gamma$ is the number of 
incoming photons, $L$ and $N_d$ are the target length and density of 
scattering centres $N_d=\frac{N_{av}\varrho}{A}$, $\Delta\Omega_n$ is the 
solid angle covered by the neutron detectors and $\varepsilon_n$ is their 
efficiency. The factor $L\Delta\Omega_n$ is estimated using a Monte Carlo 
simulation which takes into account both the real dimensions of the 
intersection of the photon beam with the target and the neutron detectors 
size and distance from the target. 
\par
While we have taken data using different $E_\gamma^{max}$ we have calculated 
the differential cross section at a given $E_\gamma$ and $\vartheta_n^{c.m.}$ 
using different sets of data, this has be done in order to have a cross check 
of the entire procedure of analysis. 
\par
\begin{figure}
\centerline{\psfig{figure=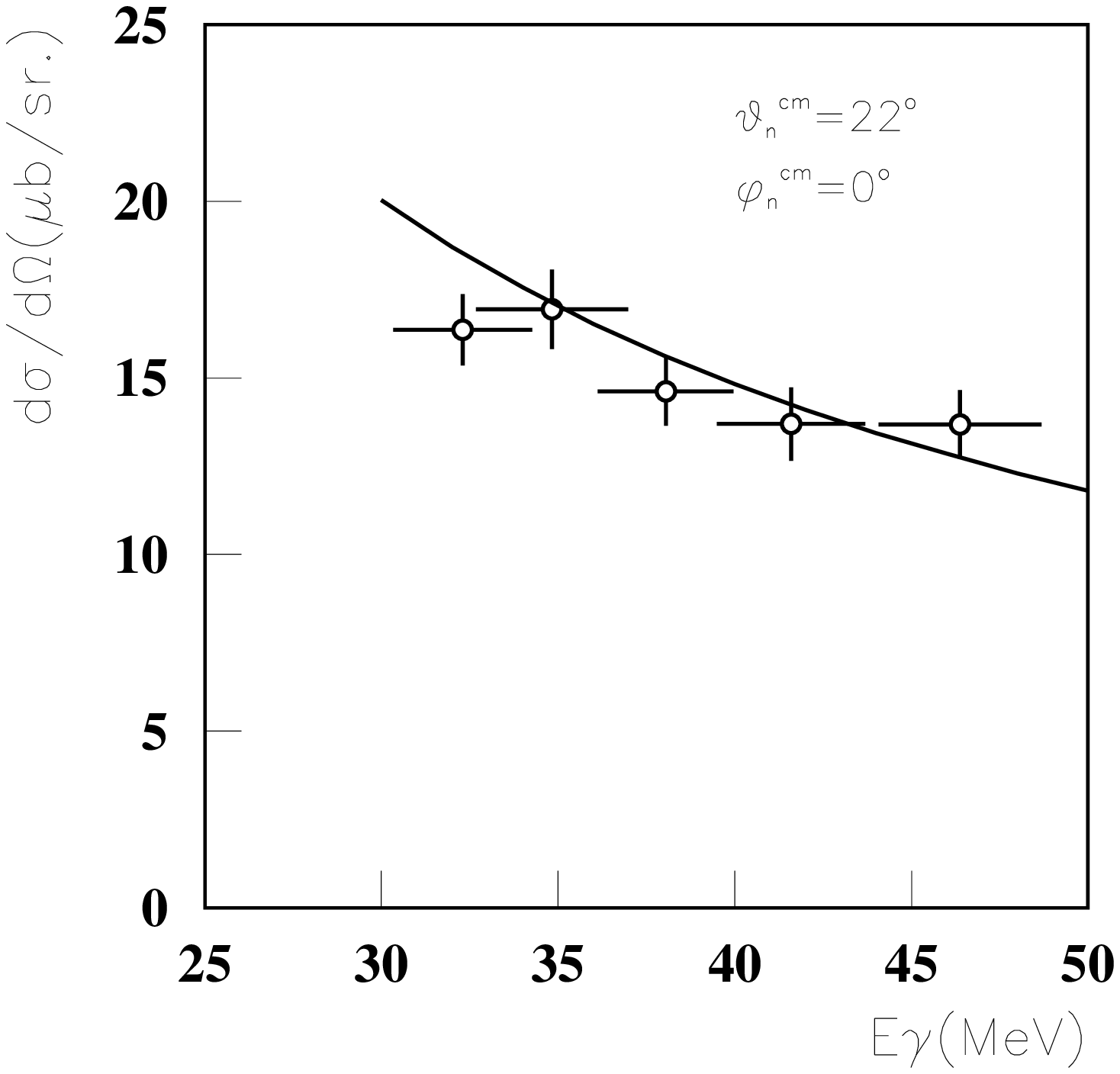,height=2.7in,width=3in}}
\centerline{\psfig{figure=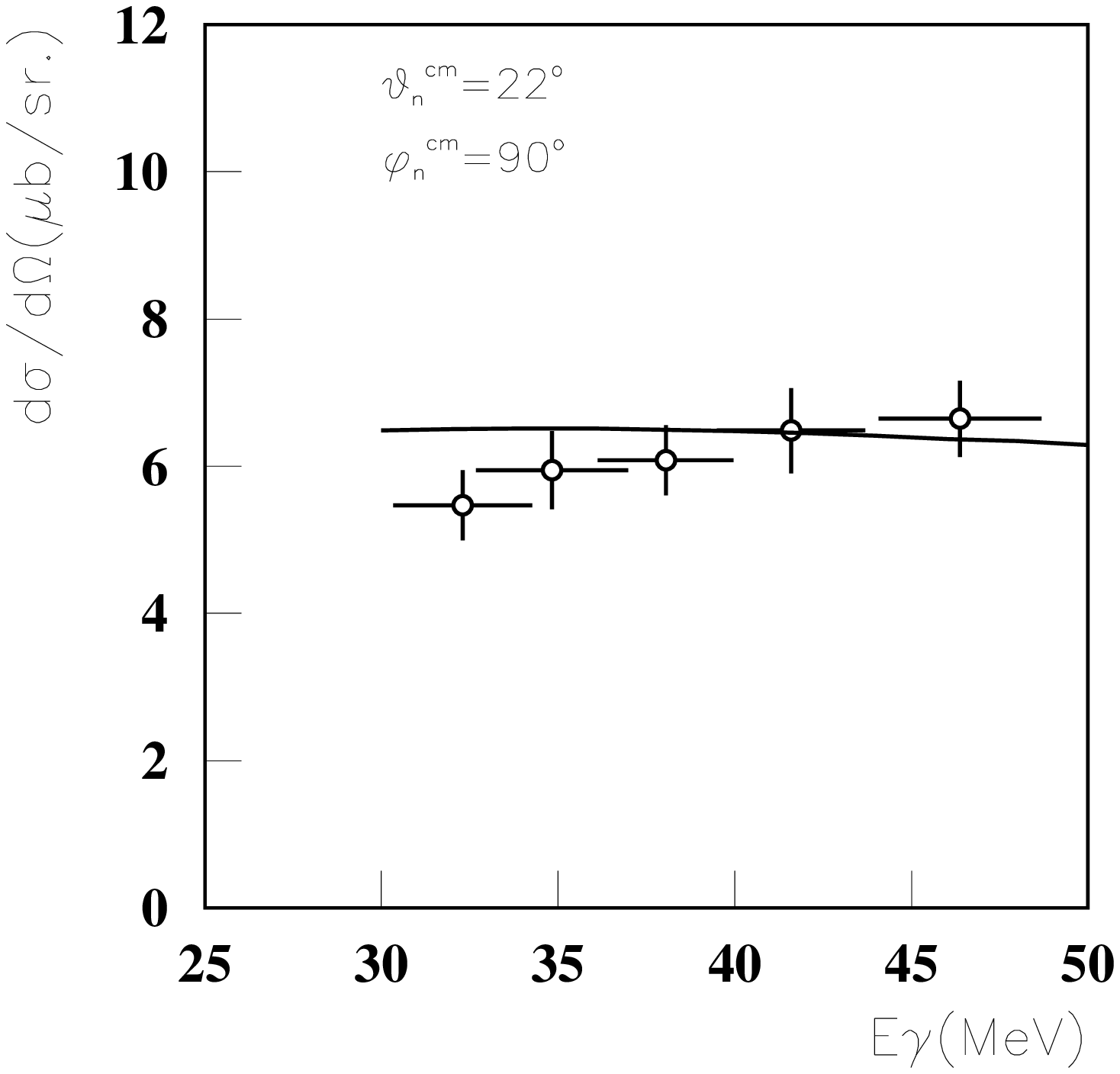,height=2.7in,width=3in}}
\centerline{\psfig{figure=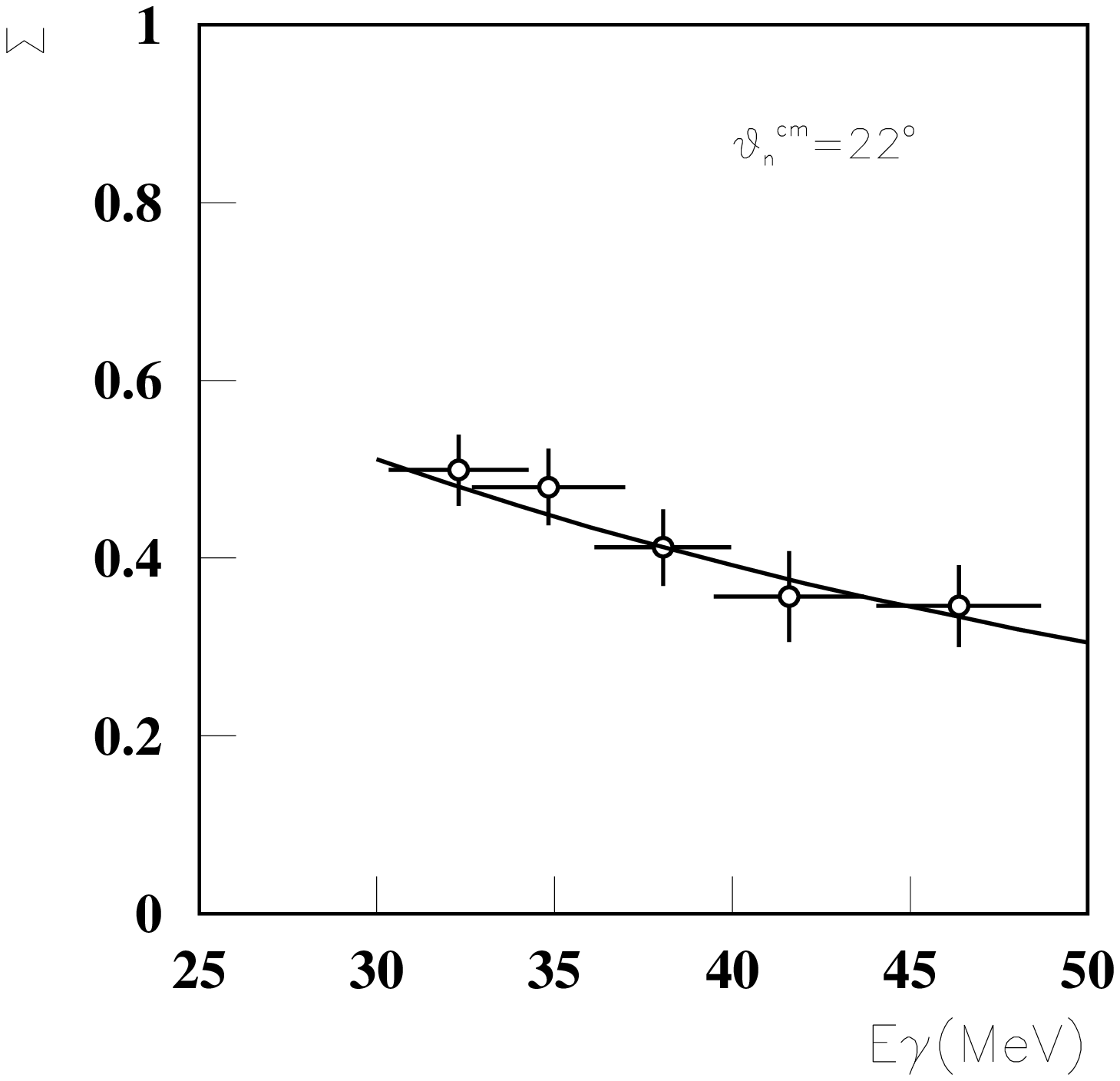,height=2.7in,width=3in}}
\caption{Parallel, Perpendicular cross section and Asymmetry for 
$\vartheta_n^{c.m.}=22^\circ$}
\label{d1}
\end{figure}
\begin{figure}
\centerline{\psfig{figure=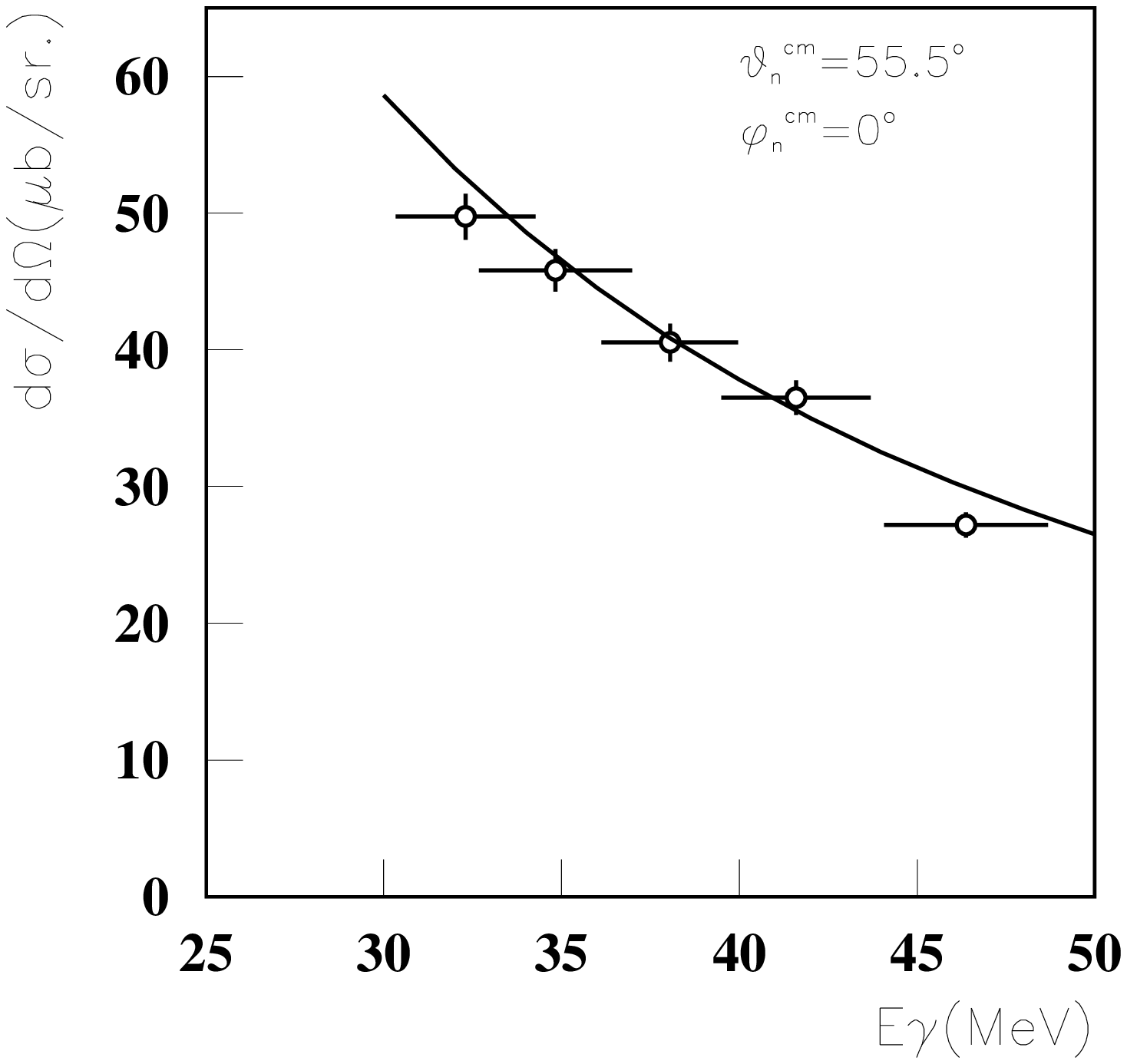,height=2.7in,width=3in}}
\centerline{\psfig{figure=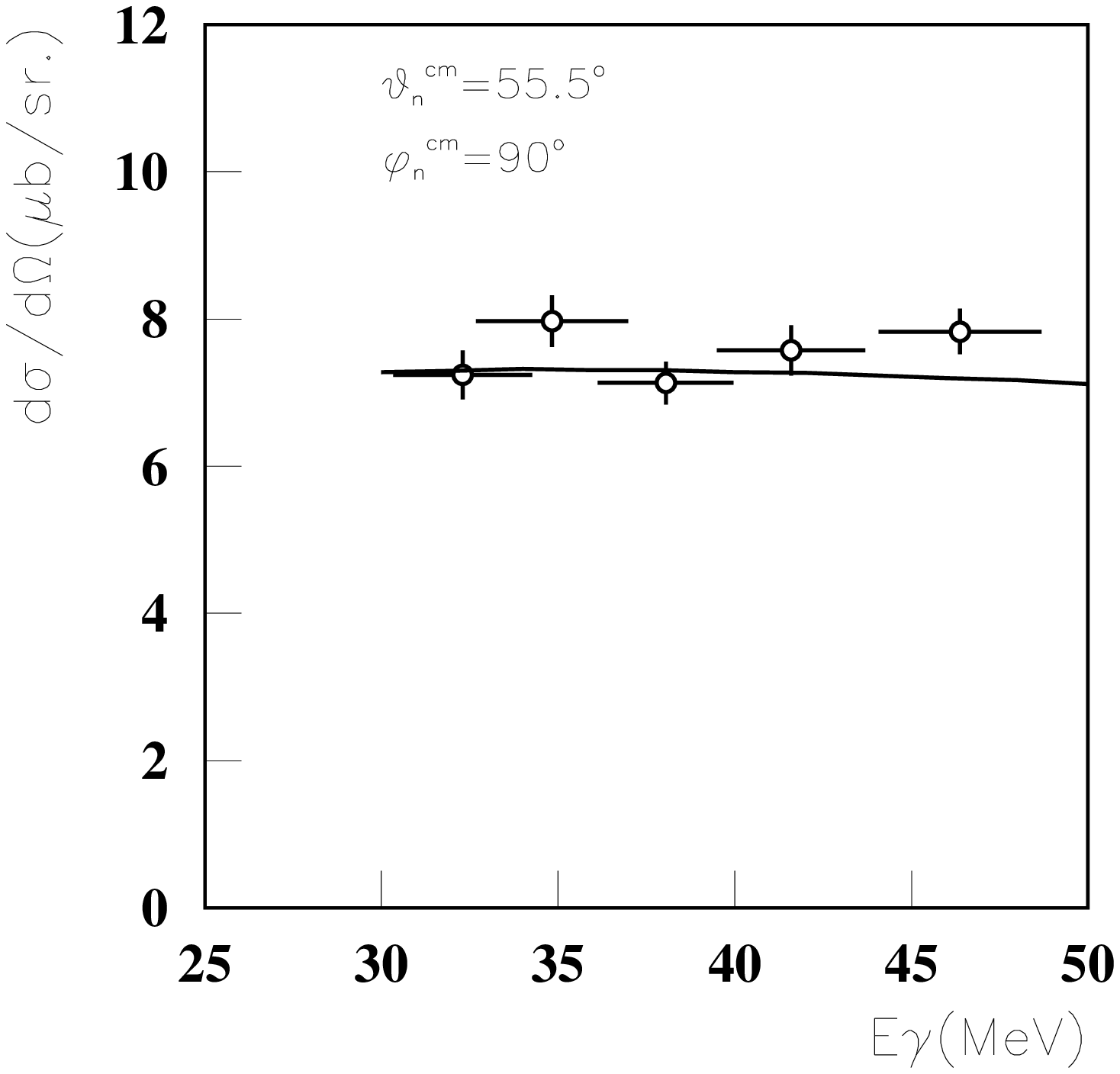,height=2.7in,width=3in}}
\centerline{\psfig{figure=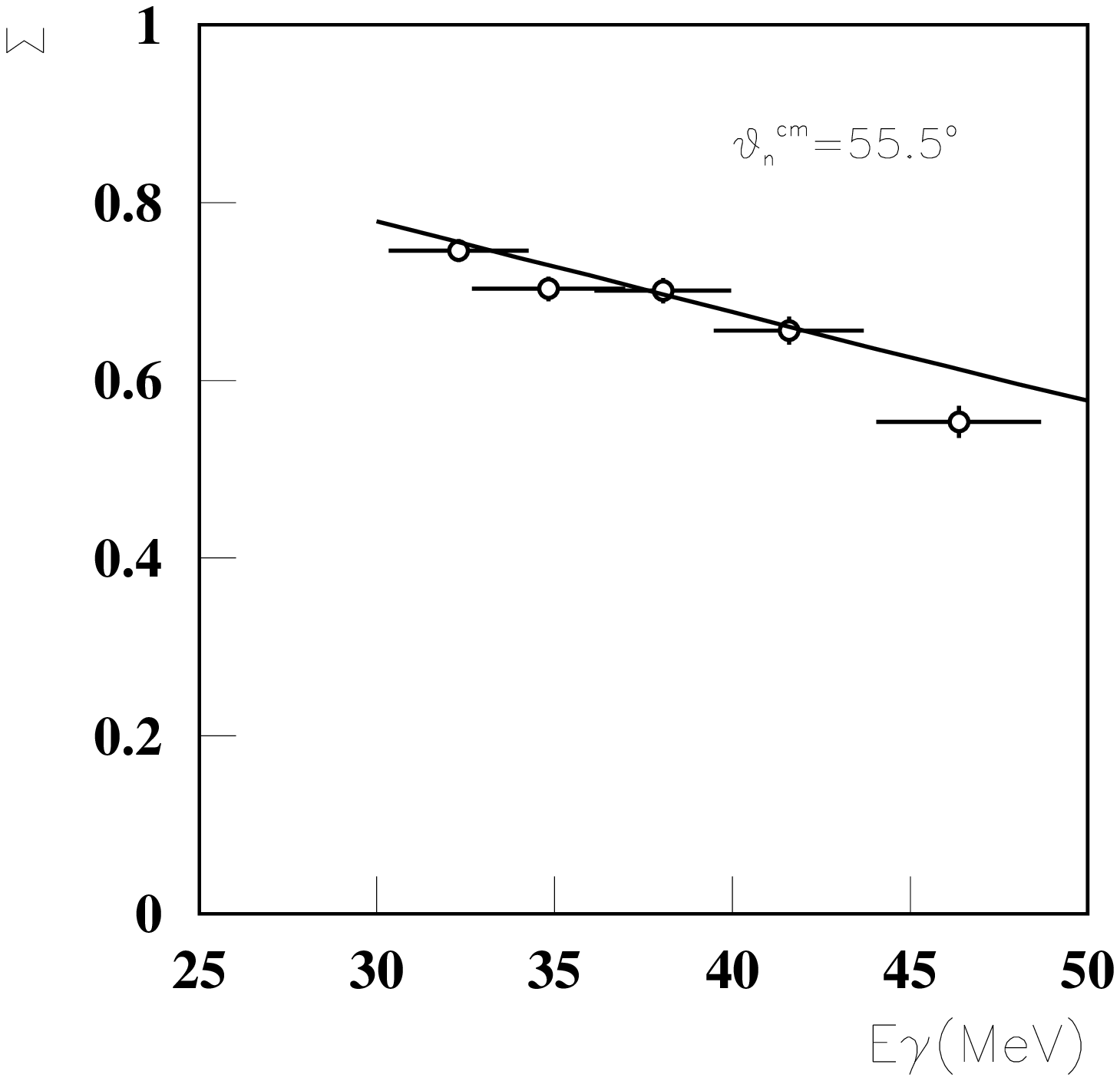,height=2.7in,width=3in}}
\caption{Parallel, Perpendicular cross section and Asymmetry for 
$\vartheta_n^{c.m.}=55.5^\circ$}
\label{d2}
\end{figure}
\begin{figure}
\centerline{\psfig{figure=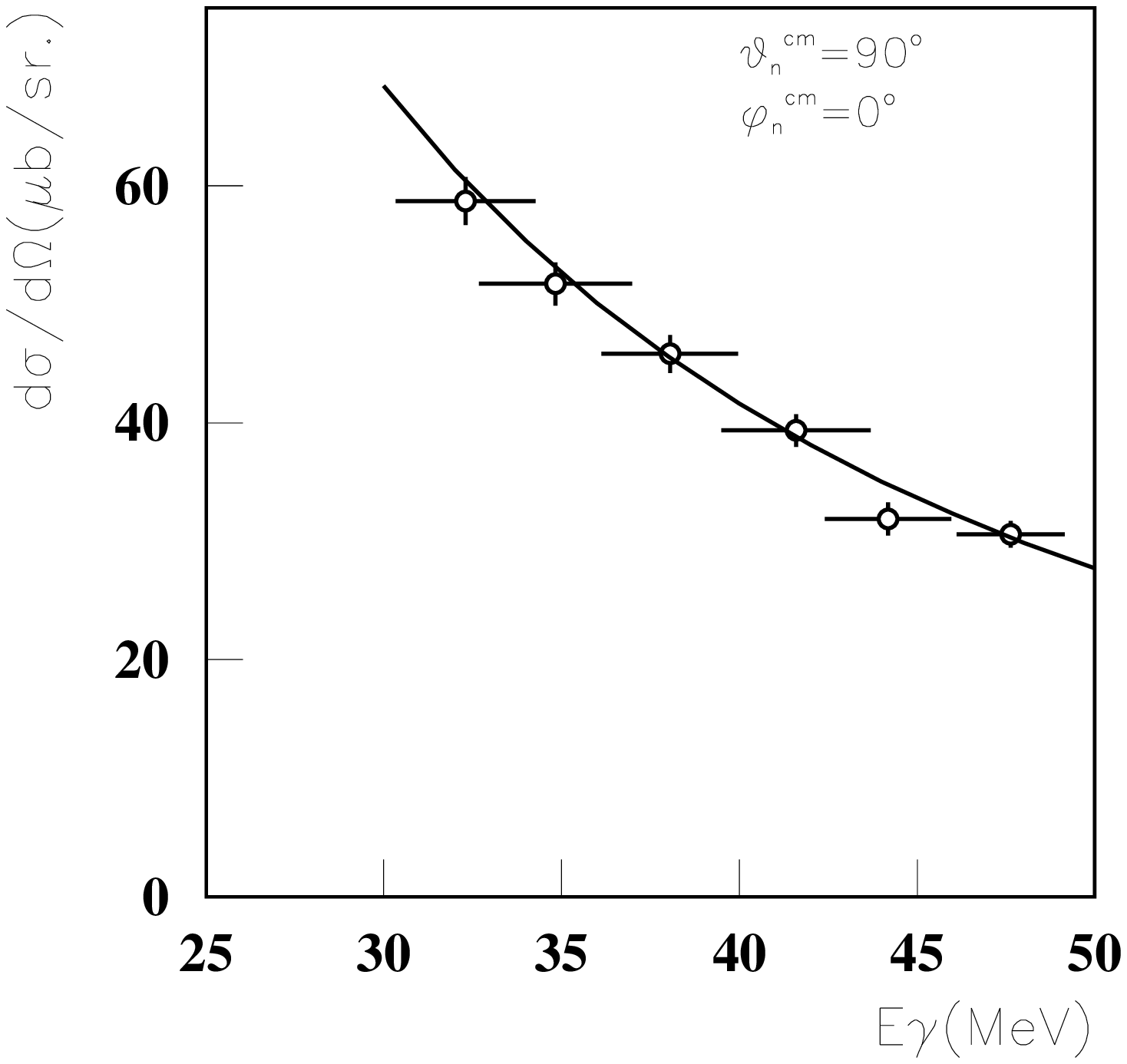,height=2.7in,width=3in}}
\centerline{\psfig{figure=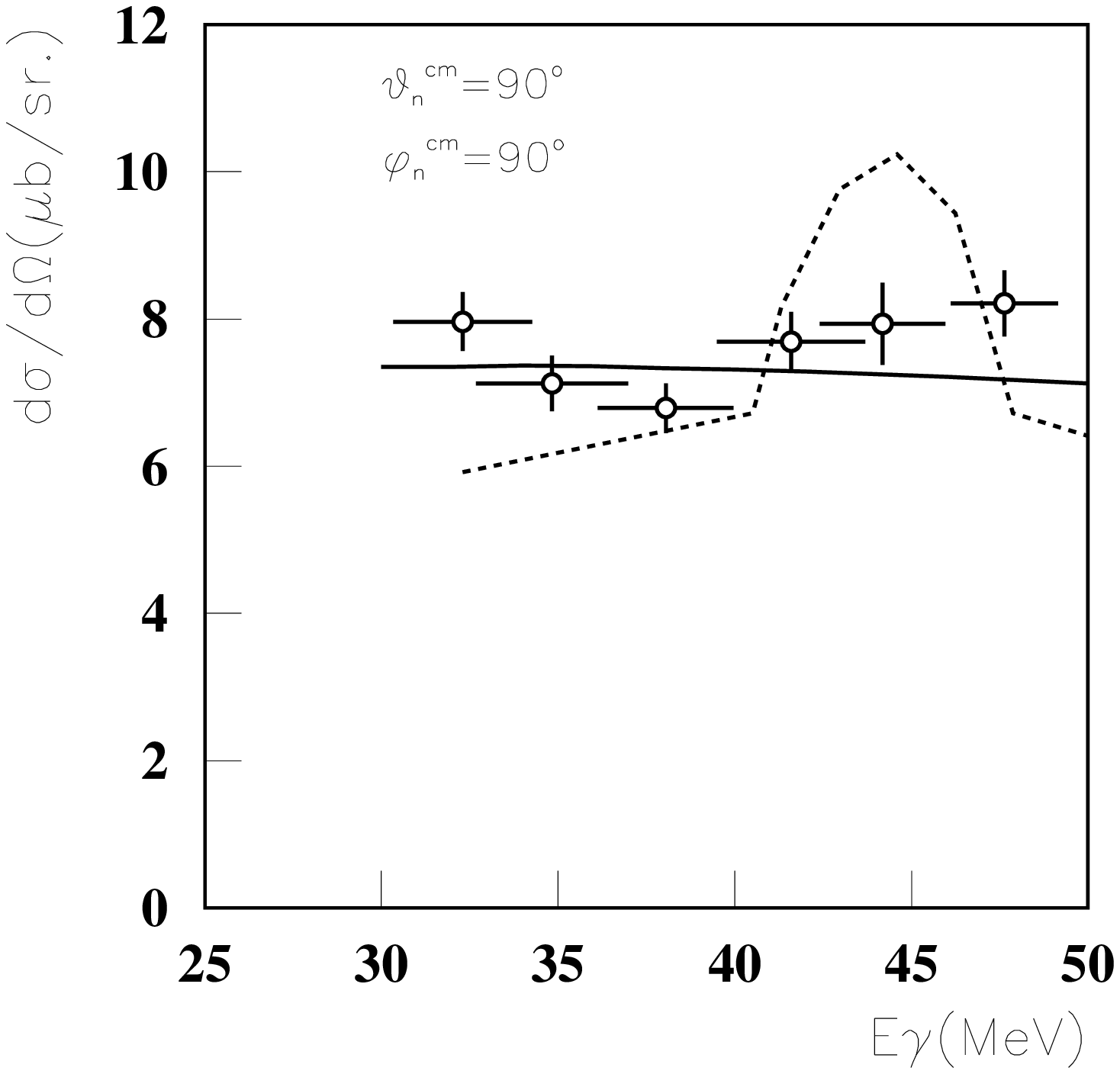,height=2.7in,width=3in}}
\centerline{\psfig{figure=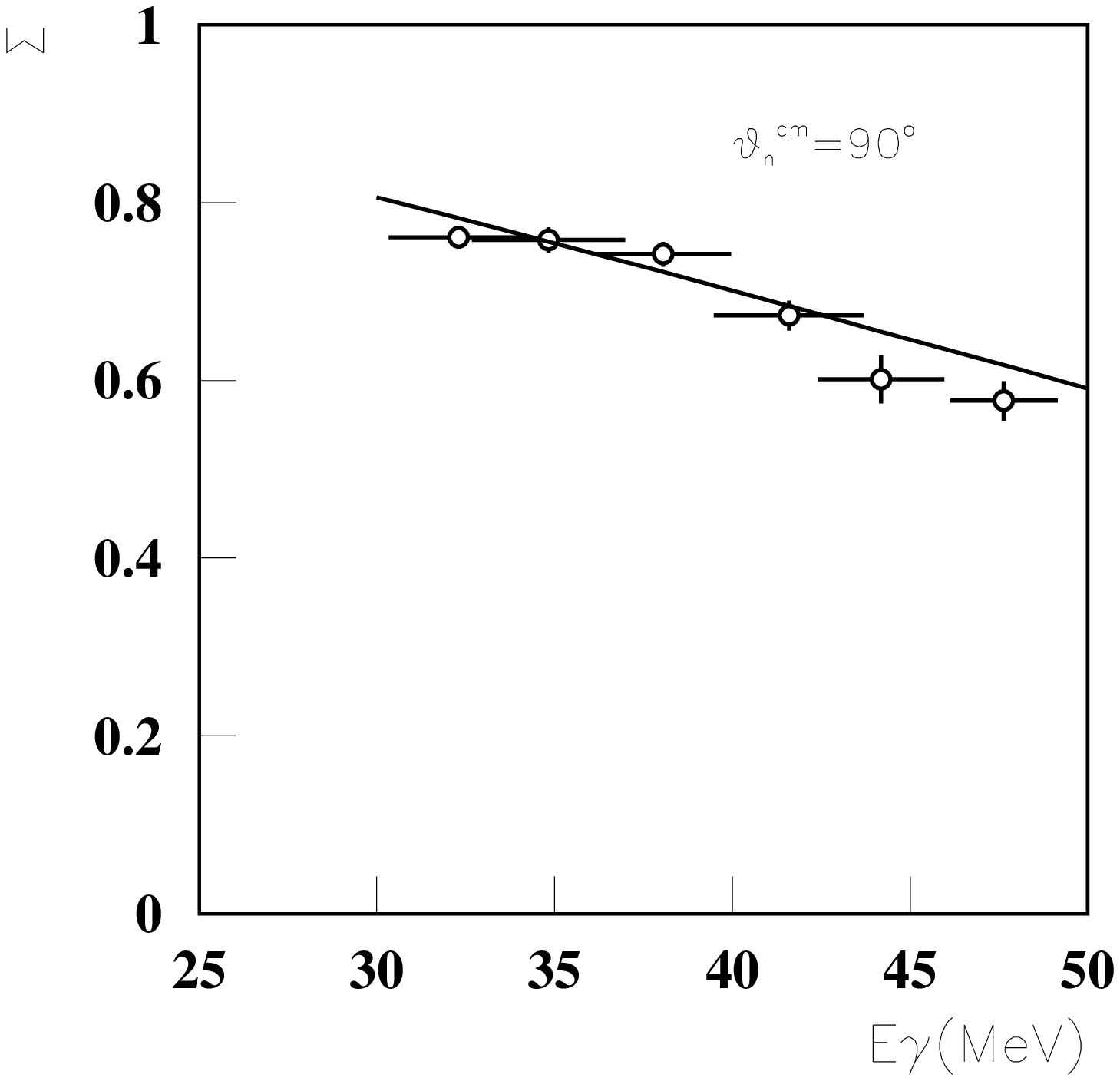,height=2.7in,width=3in}}
\caption{Parallel, Perpendicular cross section and Asymmetry for 
$\vartheta_n^{c.m.}=90^\circ$}
\label{d3}
\end{figure}
\begin{figure}
\centerline{\psfig{figure=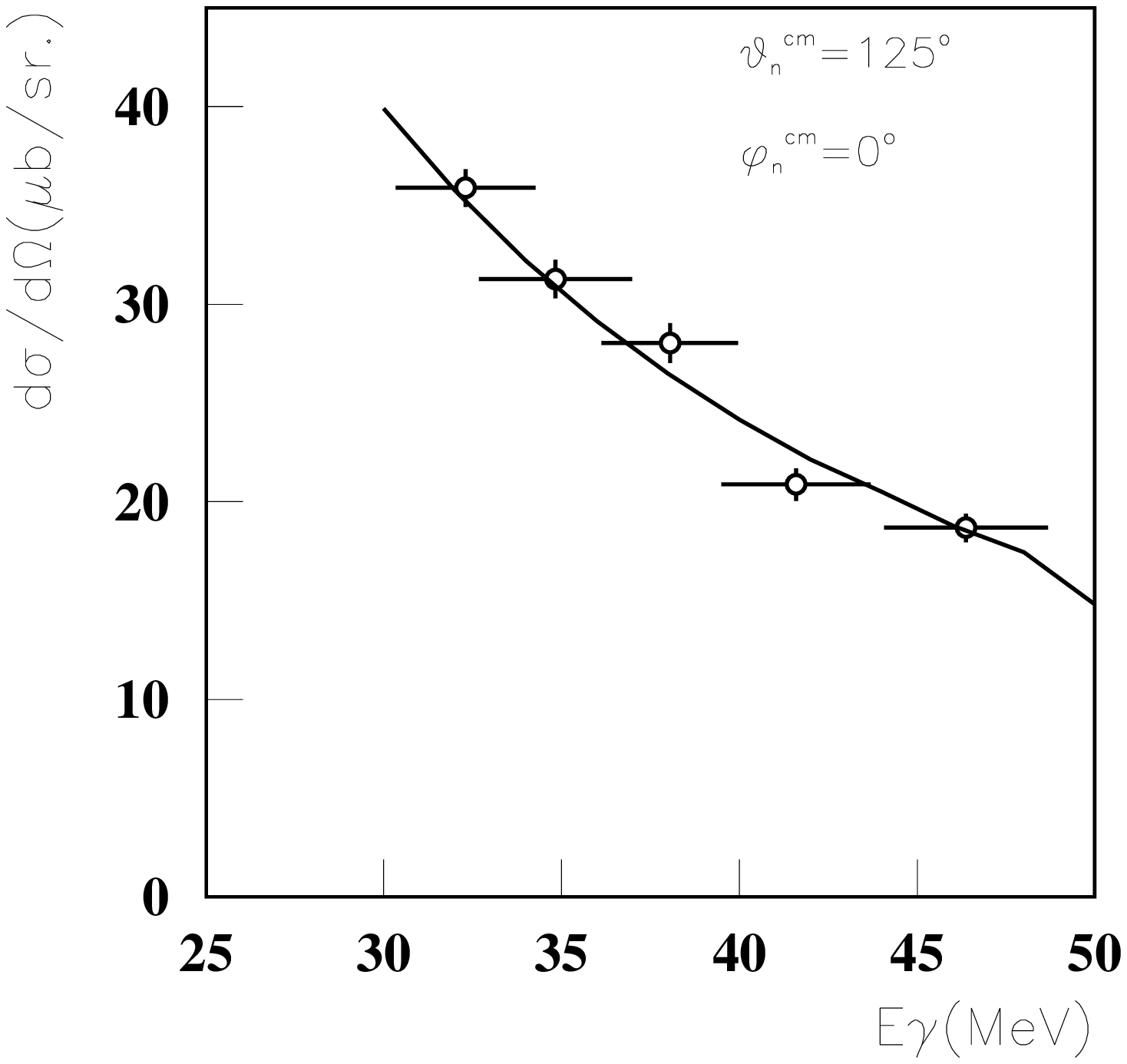,height=2.7in,width=3in}}
\centerline{\psfig{figure=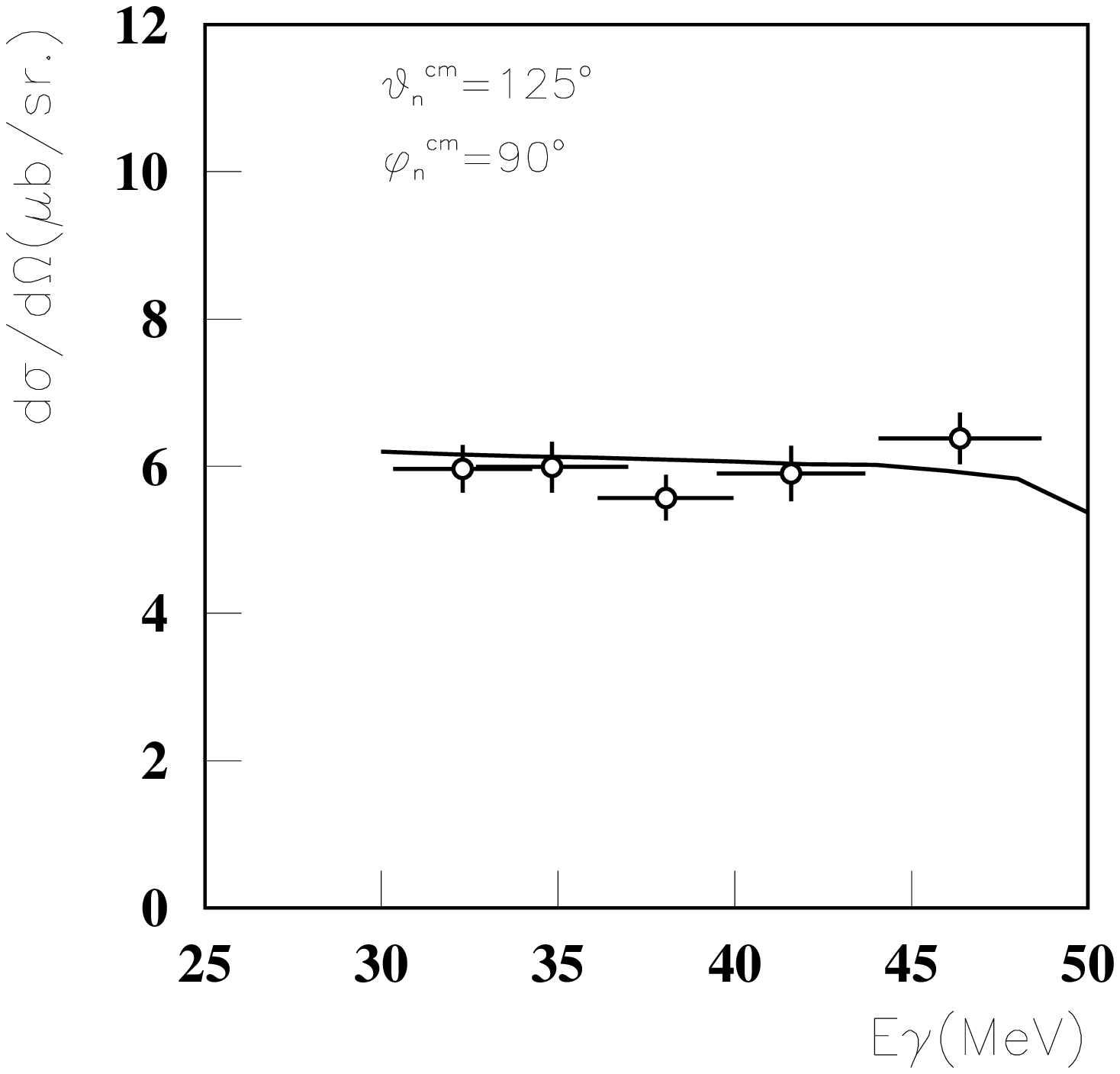,height=2.7in,width=3in}}
\centerline{\psfig{figure=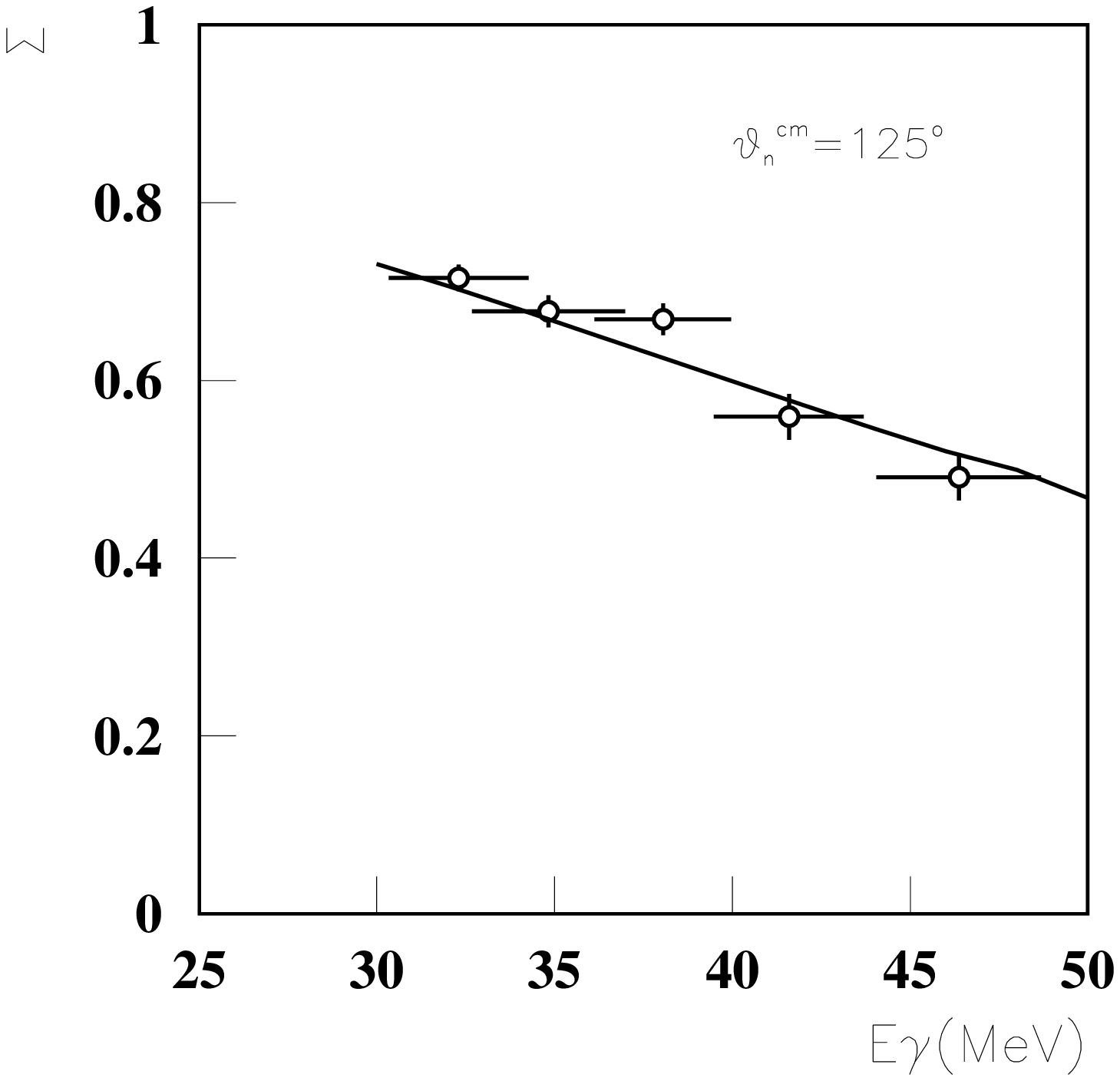,height=2.7in,width=3in}}
\caption{Parallel, Perpendicular cross section and Asymmetry for 
$\vartheta_n^{c.m.}=125^\circ$}
\label{d4}
\end{figure}
\begin{figure}
\centerline{\psfig{figure=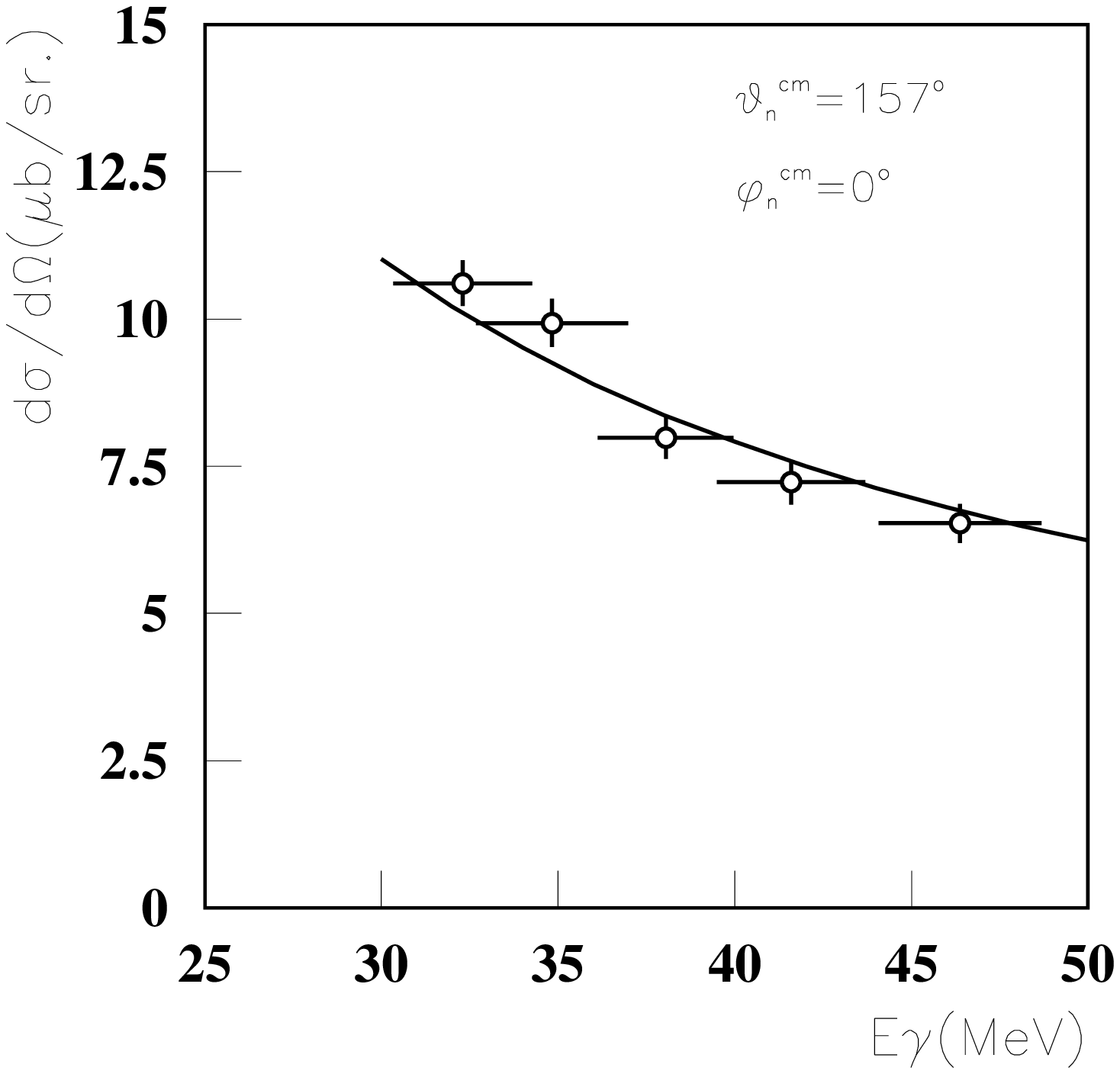,height=2.7in,width=3in}}
\centerline{\psfig{figure=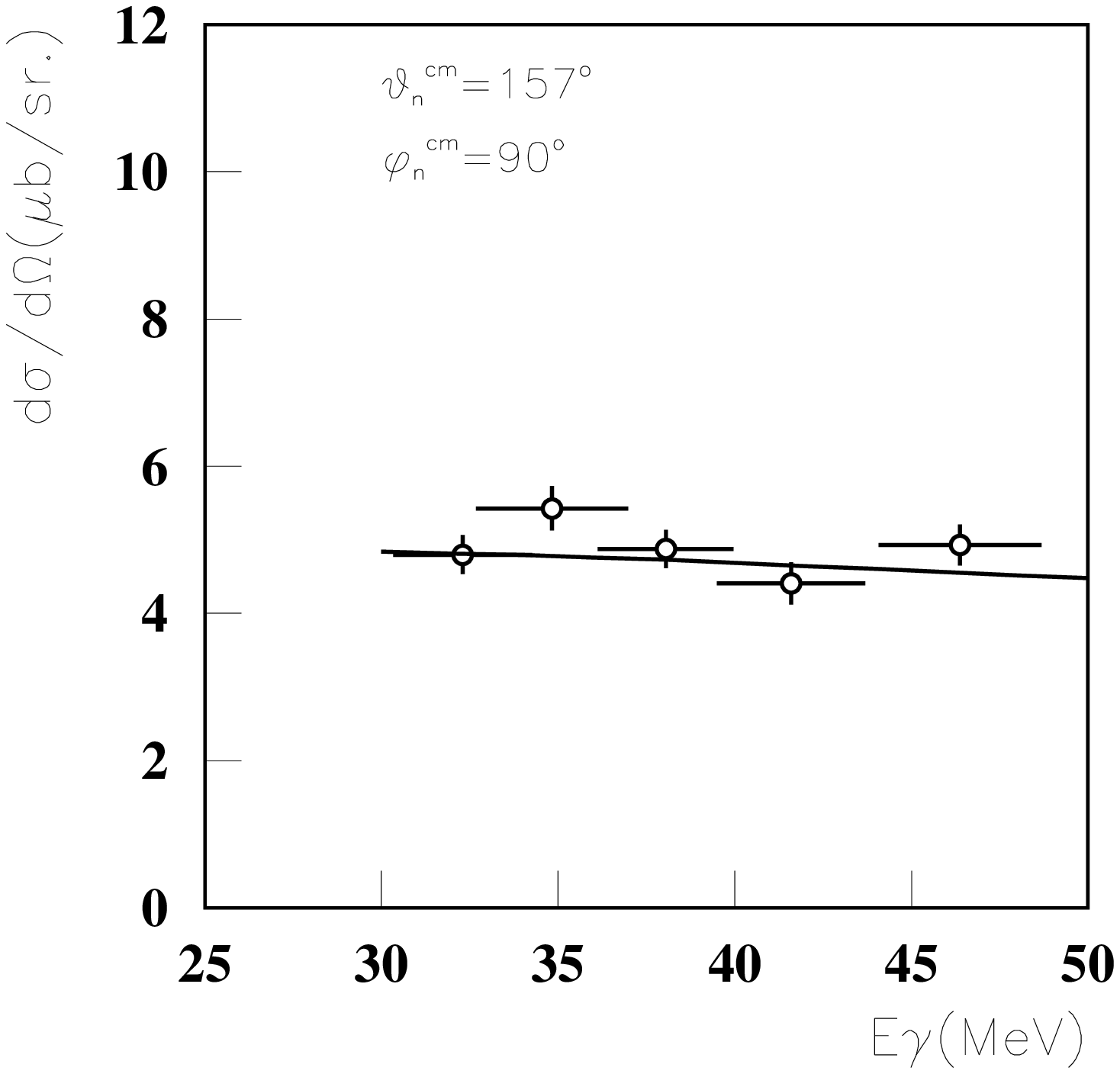,height=2.7in,width=3in}}
\centerline{\psfig{figure=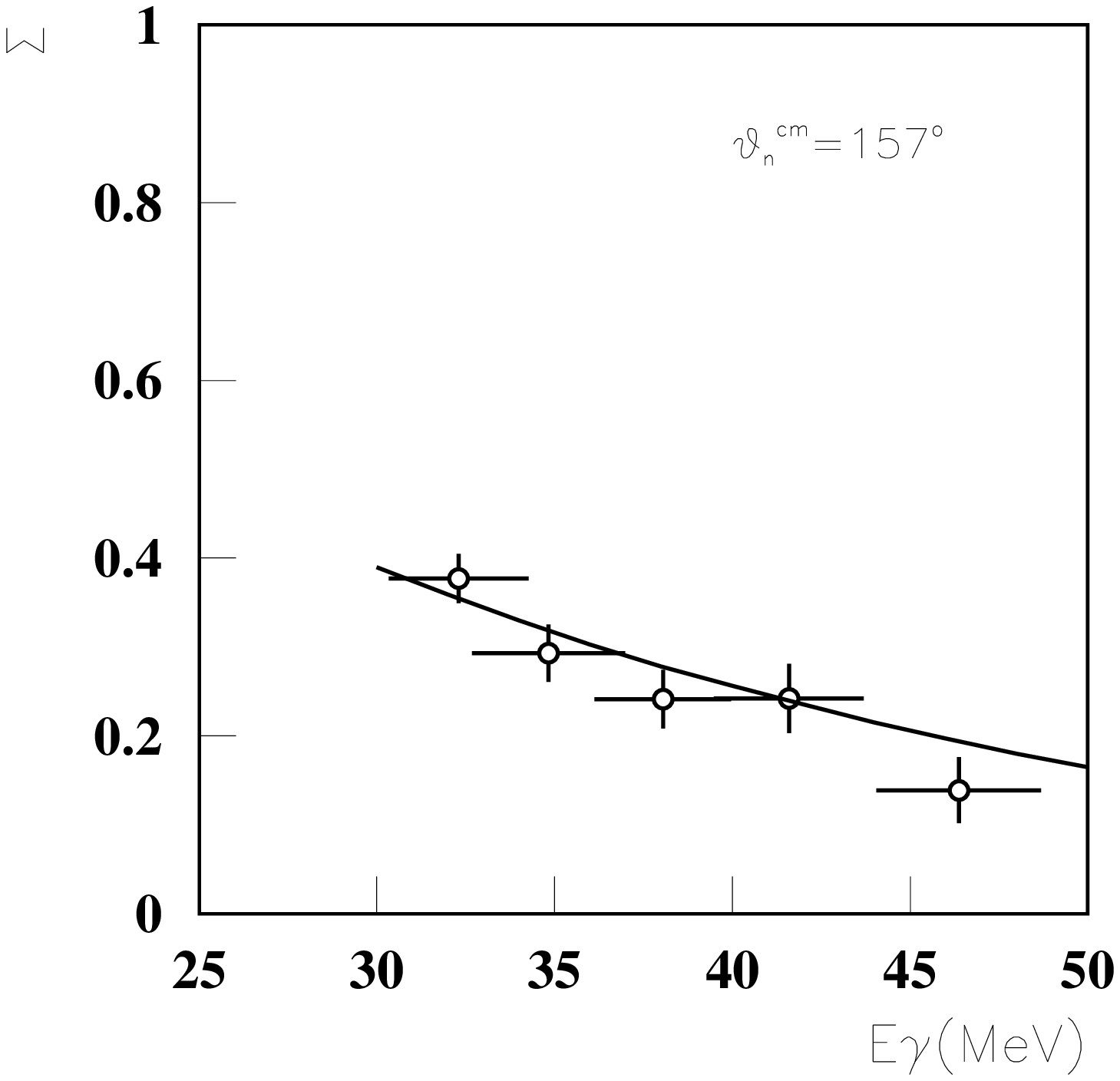,height=2.7in,width=3in}}
\caption{Parallel, Perpendicular cross section and Asymmetry for 
$\vartheta_n^{c.m.}=157^\circ$}
\label{d5}
\end{figure}

Our experimental data are shown in figures \ref{d1}-\ref{d5} where in 
each figure are illustrated the parallel cross section $d\sigma/d
\Omega$, at $\varphi_n=0^\circ$ the perpendicular cross section 
$d\sigma/d\Omega$ at $\varphi_n=90^\circ$ and the asymmetry 
$\Sigma$ for the five angles $\vartheta_n^{c.m.}$ as a function of 
the incoming $\vec\gamma-$beam energy. 
The theoretical predictions which take into account One Body Current + 
Siegert + MEC + IC + Spin Orbit Current (full lines) are from
the reference \cite{leidemann} we refer to this article for a discussion 
of the various aspects of the cross sections.
\par
In figure \ref{d3} the first proposed dibaryon resonance of \cite{russi} 
is also shown (dashed line). A fluctuation of $3.3$ standard deviation 
of the point around $44~MeV$ is necessary to make the two experiments 
compatible. The higher polarisation and lower background of the 
backscattered $\vec\gamma-$ray give us greater confidence in the quality 
of the our result.

\section{Acknowledgments}

We would like to thank W. Leidemann for the theoretical calculation 
of the cross section for the our energy and angular binning.
We also would like to thank the technical staff of the LADON facility
(E. Cima, M. Iannarelli, G. Nobili and E. Turri) for their essential
contribution in obtaining a high-quality photon beam.

\end{document}